\documentclass[12pt]{article}
 \usepackage{latexsym,amsfonts,amsmath,amsthm,amssymb}

\title{Brain as quantum-like computer}

  \author{Andrei Khrennikov\footnote{Supported in part by the EU Human
Potential Programme, contact HPRN--CT--2002--00279 (Network on
Quantum Probability and Applications) and Profile Math. Modelling
of V\"axj\"o University.}\\
International Center for Mathematical Modeling\\
in Physics and Cognitive Sciences, \\
MSI, University of V\"axj\"o, S-35195, Sweden\\
Email: Andrei.Khrennikov@msi.vxu.se}

\begin{document}
\maketitle
\date{}

\begin{abstract} We present a  contextualist statistical realistic model  for
quantum-like representations in physics, cognitive  science and psychology.
We apply this model to describe cognitive experiments
to check quantum-like structures of mental processes. The crucial role is played by interference
of probabilities for mental observables.
Recently one of such experiments based on recognition
of images was performed. This experiment confirmed our prediction on quantum-like
behaviour of mind. In our approach ``quantumness of mind'' has no direct relation to the fact that
the brain (as any physical body) is composed of quantum particles. We invented a new terminology
``quantum-like (QL) mind.'' Cognitive QL-behaviour is characterized by nonzero coefficient of interference $\lambda.$ This coefficient can be  found on the basis of
statistical data. There is predicted not only
$\cos \theta$-interference of probabilities, but also hyperbolic $\cosh \theta$-interference. This interference  was never observed for physical systems, but we could not exclude this possibility for
 cognitive systems. We propose a model of brain functioning as  QL-computer (there is discussed difference between quantum and QL computers).
\end{abstract}

 \section{Introduction}

 The idea that the description of brain functioning, cognition and consciousness could not be reduced to the theory of
neural networks and dynamical systems (cf. Ashby (1952), Hopfield (1982), Amit (1989),
Bechtel and Abrahamsen (1991), Strogatz (1994),  van Gelder (1995), van Gelder and Port (1995),
Eliasmith (1996)) and that quantum theory may play an important role in such a description  was
 discussed in the huge variety of forms, see e.g. Whitehead (1929, 1933, 1939), Orlov (1982),
 Healey (1984), Albert and Loewer (1988, 1992), Lockwood (1989, 1996), Penrose (1989, 1994),
 Donald (1990, 1995, 1996), Jibu and Yasue (1992, 1994), Bohm and Hiley (1993), Stapp (1993),
 Hameroff (1994, 1998), Loewer (1996), Hiley and Pylkk\"anen (1997), Deutsch (1997), Barrett (1999),
 Khrennikov (1999, 2000), Hiley (2000), Vitiello (2001)
and literature thereby. One of dominating approaches to application of quantum mechanics to the description of brain functioning is {\it quantum reductionism,}
see e.g. Hameroff (1994, 1998) and R. Penrose (1989, 1994). This was a new attempt of physical reduction of mental processes, cf. Ashby (1952), Hopfield (1982), Amit (1989).
This is an interesting project of great complexity and it is too early to try to make any conclusion about its future. One of important contributions of quantum reductionism in the study of mental processes is the strong critique of
the classical reductionist approach (neural networks and dynamical systems approach) and  artificial intelligence,
see especially Penrose (1994). On the other hand, quantum reductionism was strongly criticized by neurophysiologists and
cognitive scientists who still belief that neuron is the basic unit of procession of mental information. We can also
mention the {\it quantum logic} approach: mind cannot be described by classical logic and therefore the formalism
of quantum logic should be applied. It seems that Orlov (1982) published the first paper
in which this idea was explored. It is important to remark that he discussed interference within a single mind (this is natural in  quantum logic). Such an interference was also discussed by Deutsch (1997).
We pay attention to extended investigations based on {\it many-minds} approach, see Healey (1984), Albert and Loewer (1988, 1992), Lockwood (1989, 1996), Donald (1990, 1995, 1996), Loewer (1996), Barrett (1999), etc. Finally, we pay attention to attempts to apply {\it Bohmian mechanics} for description of mental processes --
Bohm and Hiley (1993), Hiley and Pylkk\"anen (1997), Hiley (2000), Khrennikov (1999, 2000), Choustova (2004).

In this paper we also develop a kind of quantum theory of mind. From the very beginning we emphasize that our approach
has nothing to do with quantum reductionism. Of course, we do not claim that our approach implies that
quantum physical reduction of mind is totally impossible. But our approach could explain the main quantum-like (QL) feature of mind -- {\it interference of minds} -- without reduction of mental processes to quantum physical processes. Regarding the quantum logic approach we can say that our contextual statistical model is quite close mathematically to some models of quantum logic (especially
Mackey's model, see e.g. Mackey (1963)), but interpretations of mathematical formalisms are totally different. The crucial point is that in our probabilistic model it is possible to combine {\it realism} with the main distinguishing features of quantum probabilistic formalism such as {\it interference of probabilities, Born's rule, complex probabilistic amplitudes,  Hilbert state space, representation of (realistic) observables by operators.}

\medskip

Why is the possibility to combine realism with quantum probabilistic features  so important for neurophysiology, cognitive sciences, psychology and sociology?

\medskip

A fundamental consequence of the possibility of such a combination is that macroscopic neuronal structures
(in particular, a single neuron) as well as cognitive and psychological contexts could exhibit quantum-like features.
Thus we may escape the fundamental problem that disturbs so much the program of the quantum physical reductionism:

\medskip

How might one combine the neuronal and quantum models?

\medskip

This was a terrible problem e.g. for Penrose (1994): {\small ``It is hard to see how one could usefully consider a quantum superposition consisting of one neuron {\it firing,} and simultaneously {\it nonfiring.}''}

\medskip

In our contextual statistical model it is possible to operate with quantum-like probabilities without such a notion as {\it superposition of states of a single
system.} All distinguishing probabilistic features of quantum mechanics can be obtained without it. This implies that
(in the opposite to quantum reductionists) we need not look for some microscopic basis of mental processes.\footnote{We remark that reductionists should do this and go to the deepest scales of space and time to find some resonable explanation of superposition and interference (S. Homeroff should go inside {\it microtubules} and R. Penrose even deeper -- to scales of {\it quantum  gravity}).} In our model {\it ``mental interference''} is not based on superposition of individual quantum states. Mental interference is described in classical (but contextual) probabilistic framework. A {\it mental wave function} represent not a mental state of an individual cognitive system, but a neurophysiological, cognitive or psychological context $C.$ \footnote{We again pay  attention that our comparation  of contextual approach and quantum reductionism could not be used as an argument against the last one. One could not exclude the possibility that  mental processes could be reduced to quantum physical processes, e.g. in microtubules, or that the act of consiousness
is really induced by the collapse of a wave function of superposition of two mass states. But our model gives the possibility to proceed with quantum mathematical formalism in neurophysiology, cognitive science, psychology and sociology without
using all those tricky things that are so important in the reductionist approach.}

As was already remarked, from the mathematical point of view our probabilistic model is quite close to the well know
Mackey's model.\footnote{In fact, Mackey's work (1963) was the starting point of my investigations and I was really lucky
that I met George Mackey at the conference of ``Quantum Structures Association'' (Castiglioncello, Italy, 1992) and discussed with him probabilistic foundations of quantum mechanics. I also was strongly influenced by Stan Gudder. His papers, see e.g. Gudder (2001), as well as numerous discussions with him played an important role in creation of my contextual statistical realistic model. General philosophic debates with C. Fuchs and A. Plotnitsky, see even
Fuchs (2002) and Plotnitsky (2002), played an  important role in creation of so called V\"axj\"o interpretation, see Khrennikov (2002).} George Mackey (1963) presented  a program of huge complexity and importance:

\medskip {\it To deduce the probabilistic formalism of quantum mechanics starting with a system of natural probabilistic
axioms.}

\medskip

(Here ``natural'' has the meaning a natural formulation in classical probabilistic terms.) G. Mackey tried to realize this program starting with a system of  8 axioms -- Mackey axioms, see Mackey (1963). This was an important step in clarification
of the probabilistic structure of quantum mechanics. However, he did not totally succeed (as was recognized by himself, see Mackey (1963) and section 2 for details). The crucial axiom (about the complex Hilbert space) was not formulated in natural
(classical) probabilistic terms.

In Khrennikov (2001, 2002, 2003) there was  presented a new attempt of realization of Mackey's program. In our approach the probabilistic structure of quantum mechanics (including the complex Hilbert space) can be derived on the basis of two axioms formulated in classical
(but contextual!) probabilistic terms. This realization of Mackey's program gives the possibilty to combine
realism and quantum probabilistic behavior (see previous discussion on quantum reductionism of mental prosses).

Comparing the Bohmian mental models and our contextual quantum-like model we can say that our model does not provide individual description of mental processes. We could not describe a ``trajectory of an individual
mind'', we are able only describe ``probability distributions of minds.'' By using the terminology of Atmanspacher et al. (2001) we can say that Bohmian mental models provide the ontic description of mental processes and our model
provides the epistemic description. Our epistemic (contextual probabilistic) model does not contradict to the possibility
that on the ontic level mental world can be described by Bohmian mechanics. On the other hand, our model does not imply that precisely Bohmian mechanics is  the right ontic mental model. In principle, there might be found other ontic mental models
which could be more adequate the problem under consideration. Another fundamental feature of Bohmian mechanics is its
{\it nonlocality.} Nonlocality of this model has important consequences for cognitive science; especially the problem of consciousness. Bohm, Hiley and Pylkk\"anen developed a new philosophic systems -- philosophy of wholeness and applied
it to the problem of consiousness. Khrennikov (1999, 2000) proposed a mathematical model of {\it mental space} given by an infinite $p$-adic tree. Minds were represented by infinite branches of tree; they were coupled by mental pilot
waves. In Choustova (2004) there was presented a model of psycho-financial market with traders coupled by mental pilot
waves.

We say a few words about many-minds approach: Healey (1984), Albert and Loewer (1988, 1992), Lockwood (1989, 1996), Donald (1990, 1995, 1996), Loewer (1996), Barrett (1999), etc.. This approach played an important role in justification of the many-worlds interpretation of quantum mechanics. There is no direct contradiction between our contextual realistic and  many-worlds interpretations. We recall that the many-worlds interpretation
was invented as an attempt to explain some mysteries of quantum mechanics. We agree that there were naturally explained a few  things that
were really mysterious in the orthodox Copenhagen interpretation.\footnote{This interpretation is typically considered in the quantum reductionist approach.}  In particular, it seems that only the many-worlds interpretation provides a resonable exlpanation
of {\it quantum parallelism} (which plays the fundamental role in quantum computing). Therefore many-minds interpretation  is
so natural for various models of brain as a physical device performing quantum computations. We also can escape conventional
Copenhagen mysteries by using our contextual realistic model.  Connections between our model and the many-worlds/many-minds approaches will be discussed in more detail in section 12.

Before having a closer look at our model, it is perhaps necessary to
discuss the meaning of the term {\it contextuality}, as it can
obviously be interpreted in many different ways. The
most common meaning (especially in the literature on quantum logic)
is that the outcome for a measurement of an observable $u$ under
a contextual model is calculated using a different (albeit hidden)
measure space, depending on whether or not compatible observables
$v, w,...$ were also made in the same experiment. We remark
that well known ``no-go'' theorems cannot be applied to such
contextual models.\footnote{This approach to contextuality can be considered as a mathematical
formalization of {\it Bohr's measurement contextuality,} see Plotnitsky (2001, 2002) for details.
Bohr's interpretation of quantum mechanics is in fact considered as
contextual;  for N.
Bohr the word ``context'' had the meaning of a {\it ``context of a
measurement.''}}

In our approach  the term contextuality is used in
a totally different meaning. Roughly speaking our approach is
noncontextual from the conventional viewpoint. Values
associated to  two specially chosen observables
 -- the {\it reference observables} \footnote{For example, physical observables,
or observables on neuronal structures, or observables corresponding to performing of cognitive or psychological
tasks, or questions asked to a group of people. E.g., position and
momentum in physics; in Conte et al. (2004) there were considered two cognitive tasks based on recognition of hidden structures of two different
images.} -- are considered as objective properties of
systems (physical, cognitive,..., social). These observables
are therefore not contextual in the sense of Bohr's measurement
contextuality.

The basic notion of our approach is the {\it context} -- that is, a
complex of physical, biological, cognitive, psychological, social, or economic
conditions. Systems (physical, biological, ..., economic) interact with a
context $C$ and in this process a statistical ensemble $S_C$
is formed; cf. e.g. Ballentine (2001).\footnote{The notion of context is
close to the notion of preparation procedure, see e.g. Holevo (2001).
However, for any preparation procedure ${\cal
E},$ it is assumed that this procedure could be (at least in
principle) realized experimentally. We do not assume this for an
arbitrary context $C.$ }

 Conditional (or better to say contextual)
probabilities  for reference observables, ${\bf P}(a=y/C), {\bf
P}(b=x/C),$ are used to represent the context $C$ by a complex
probability amplitude $\psi_C.$ This amplitude is in fact encoded in
a {\it generalization of the formula of total probability describing
the interference of probabilities.} Note that interferences of
probabilities can thus be obtained in a classical probabilistic
framework (i.e., without the need of the Hilbert space formalism),
an observation which was actually the starting point of our
considerations. Our approach is thus based on two cornerstones:

\medskip

a) {\it contextuality of probabilities};

\medskip

b) {\it the use of two fixed incompatible observables (physical, biological, cognitive, psychological, social, or economic)  in order to represent the classical contextual probabilistic model in
the complex Hilbert space.}

\medskip

Section 2.1 is devoted to the presentation of our general contextual statistical model (V\"axj\"o model); in section 2.2.
there is considered the ensemble representation of contextual statistical models (cf. with so called statistical interpretation of quantum mechanics of Einstein, Margenau, ..., see e.g. Ballentine (2001), Holevo (2001);
in section 2.3 we discuss the possibility to apply this model outside physics (in particular, in cognitive science,
psychology and sociology). In sections 4--10 we apply our model to description of mental observations in the quantum-like terms. We start with {\it mental interference} which is defined as interference of probability distributions of two
{\it incompatible mental observables.} For example, in psychology such observables can be realized in the form of two
incompatible questions which are asked to people participating in a test.

In our model {\it incompatibility of two mental
observables} is defined in purely classical probabilistic terms.\footnote{This classical probabilistic incompatibility
implies noncommutativity of operators $\hat{a}$ and $\hat{b}$ corresponding to mental observables $a$ and $b.$ But such
a representation is not basic; it is induced by classical (contextual) proabbilistic representation.} A condition of incompatibility can be easily checked on the basis of experimental statistical data collected e.g. in the form of
``yes-no'' answers to questions. The magnitude of mental interference is characterized by a coefficient of interference
(or incompatibility) $\lambda.$ Depending on this magnitude we obtain different representations of probabilities
in experiments with cognitive systems. In particular, we obtain the quantum-like representation in the complex Hilbert space.

This approach should be justified experimentally. A priory there are no reasons that cognitive systems may exhibit the quantum-like probabilistic behaviour; in particular, nontrivial mental interference. Therefore we presented
the detailed description of an experimental test  to check the hypothesis on the quantum-like probabilistic behaviour. We hope that such tests would be performed in various domains of mental sciences: psychology, cognitive science,
sociology, economics. Some preliminary experiments were already done Conte et al. (2004) and they confirmed that in some psychological experiments students can exhibit quantum-like probabilistic behaviour.
In section 8 we present a model of brain's functioning as QL-computer. There is discussed  difference between conventional quantum and QL computers, especially regarding to quantum parallelism.
In section 12 we shall compare in more detail
the many-minds and V\"axj\"o approaches. Finally, in section 13 we discuss the notion of
{\it macroscopic quantum system} and give motivations to consider human beings as macroscopic quantum systems.

The paper is written to be readable by researchers working in biology, cognitive and social sciences, psychology.
There is used not so much mathematics. Only in section 5 we present rather long mathematical expressions. In principle,
this section can be omitted if one accepts that there exists an algorithm which gives the possibility to construct
a complex probability amplitude -- ``wave function'' -- on the basis of contextual probabilities.

 \section{Contextual statistical realistic model}
A general statistical realistic model for observables based on the
contextual viewpoint to probability will be presented. It
will be shown that classical as well as quantum probabilistic models
can be obtained as particular cases of our general contextual
model, the {\it{V\"axj\"o model}}.\footnote{
    This model is not reduced to the conventional, classical and quantum
    models. In particular, it contains a new statistical model: a model with
hyperbolic $cosh$-interference that induces  "hyperbolic quantum
mechanics", Khrennikov (2003).} Realism is one of the main distinguishing features of
the V\"axj\"o model since it is always possible manipulate
objective properties, despite the presence of such essentially
quantum effects as, e.g., the interference of probabilities.

As George W. Mackey (1963) pointed out, probabilities cannot be
considered as abstract quantities defined outside any reference to a
concrete complex of physical conditions $C.$ All probabilities are
conditional or better to say contextual.\footnote{We remark
that the same point of view can be found in the works of A. N.
Kolmogorov and R. von Mises. However, it seems that Mackey's
book was the first thorough presentation of a program of conditional
probabilistic description of measurements, both in classical and
quantum physics.} G. Mackey did a lot to unify classical and quantum
probabilistic description and, in particular, demystify quantum
probability. One crucial step is however missing in Mackey's work.
In his book Mackey (1963) introduced the quantum probabilistic model
(based on the complex Hilbert space) by means of a special axiom
(Axiom 7, p. 71) that looked rather artificial in his
general conditional probabilistic framework.

Mackey's model is based on a system of eight axioms, when our own
model requires only two axioms. Let us briefly mention the content
of Mackey first axioms. The first four axioms concern conditional
structure of probabilities, that is, they can be considered as
axioms of a classical probabilistic model. The fifth and sixth
axioms are of a logical nature
    (about questions).
We reproduce below Mackey's ``quantum axiom'', and Mackey's own
comments on this axiom (see  pp. 71-72):

\medskip

{\bf Axiom 7} (G. Mackey) {\it The partially ordered set of all questions in quantum
mechanics is isomorphic to the partially ordered set of all closed
subsets of a separable, infinite dimensional Hilbert space.}\footnote{``This axiom has rather a different character from Axioms 1
through 4. These all had some degree of physical naturalness and
plausibility. Axiom 7 seems entirely {\it ad.hoc.} Why do we make
it? Can we justify making it? What else might we assume? We shall
discuss these questions in turn. The first is the easiest to answer.
We make it because it ``works'', that is, it leads to a theory
which explains physical phenomena and successfully predicts the
results of experiments. It is conceivable that a quite different
assumption would do likewise but this is a possibility that no one
seems to have explored. Ideally one would like to have a list of physically
plausible assumptions from which one could deduce Axiom 7.''}

\medskip

Our activity can be considered as an attempt to find a list of
physically plausible assumptions from which the Hilbert space
structure can be deduced. We show that this list can consist in two
axioms (see our Axioms 1 and 2) and that these axioms can be
formulated in the same classical probabilistic manner as Mackey's
Axioms 1--4.

\subsection{Contextual statistical model of observations}
A physical or mental {\it context}  $C$ is  a complex of physical or mental conditions.
Contexts are fundamental elements of any contextual statistical model. Thus construction of any model
$M$ should be started with fixing the collection of  contexts of this model; denote the collection of contexts
by the symbol ${\cal C}$ (so the family of contexts  ${\cal C}$ is determined by $ M).$ In mathematical formalism ${\cal C}$ is an abstract set
(of ``labels'' of contexts). Another fundamental element of any contextual statistical model $ M$ is a set of observables ${\cal O}:$
any observable $a\in {\cal O}$ can be measured
under a complex of physical conditions $C\in {\cal C}.$  For an $a \in {\cal O},$ we denote the set of its possible values (``spectrum'') by the symbol
$X_a.$

We do not assume that all these observables can be measured simultaneously. To simplify considerations, we shall consider only discrete observables
and, moreover, all concrete investigations will be performed for {\it dichotomous observables.}

\medskip

{\bf Axiom 1:} {\it For  any observable
$a \in {\cal O},$ there are defined contexts $C_\alpha$
corresponding to $\alpha$-filtrations: if we perform a measurement of $a$ under
the complex of physical conditions $C_\alpha,$ then we obtain the value $a=\alpha$ with
probability 1. It is supposed that the set of contexts ${\cal C}$ contains filtration-contexts $C_\alpha$
for all observables $a\in {\cal O}.$}

\medskip

{\bf Axiom 2:} {\it There are defined contextual probabilities ${\bf P}(a=\alpha/C)$ for any
context $C \in {\cal C}$ and any observable $a \in {\it O}.$}

\medskip

Probabilities ${\bf P}(b=\beta/C)$ are interpreted as {\it contextual (conditional) probabilities.} Especially important role will be played by probabilities:
$$
p^{a/b}(\alpha/\beta)\equiv {\bf P}(a=\alpha/C_\beta), a, b \in {\cal O}, \alpha \in X_a, \beta \in X_b,
$$
where $C_\beta$ is the $[b=\beta]$-filtration context. For any $C\in {\cal C},$ there is defined the set of probabilities:
$
 \{ {\bf P}(a=\alpha/C): a \in {\cal O}\}.
$
We complete this probabilistic data by $C_\beta$-contextual probabilities:
$$
D({\cal O}, C)= \{ {\bf P}(a=\alpha/C),{\bf P}(b=\beta/C), ...,
{\bf P}(a=\alpha/C_\beta), {\bf P}(b=\beta/C_\alpha),...\},
$$
where $a,b,... \in {\cal O}.$
We denote  the collection of probabilistic data
$D({\cal O}, C)$ for all contexts  $C\in {\cal C}$ by the symbol
${\cal D}({\cal O}, {\cal C}).$\footnote{We remark that $D({\cal O}, C)$ does not contain the simultaneous probability distribution of
observables $a, b \in \cal O$ (under the context $C).$
Data $D({\cal O}, C)$ gives a probabilistic image of the context $C$ through the
system of observables ${\cal O}.$
There is defined the map:
$\pi :{\cal C} \to {\cal D}({\cal O}, {\cal C}), \; \; \pi(C)= D({\cal O}, C).$
In general this map is not one-to-one. Thus the $\pi$-image of contextual reality is very rough:
{\it not all contexts can be distinguished with the aid of probabilistic data produced by the class
of observables ${\cal O}.$ }}

\medskip

{\bf Definition 2.1.} {\it A contextual  statistical model of reality is a triple
\begin{equation}
\label{VM}M =({\cal C}, {\cal O}, {\cal D}({\cal O}, {\cal C}))
\end{equation}
where ${\cal C}$ is a set of contexts and ${\cal O}$ is a  set of observables
which satisfy to axioms 1,2, and ${\cal D}({\cal O}, {\cal C})$ is probabilistic data
about contexts ${\cal C}$ obtained with the aid of observables ${\cal O}.$}

\medskip

We call observables belonging to the set ${\cal O}\equiv {\cal O}(M)$ {\it reference of observables.}
Inside of a model $M$  observables  belonging  to the set ${\cal O}$ give the only possible references
about a context $C\in {\cal C}.$

\medskip

{\bf Definition 2.3.} {\it Reference observables are  said to be mutually incompatible if
\begin{equation}
\label{VM5}
p^{a/b}(\alpha/\beta)\not= 0, \alpha \in X_a, \beta\in X_b,
\end{equation}
for any pair $a,b \in {\cal O}.$}

\medskip

We shall see that in the case ${\cal O}= \{ a, b\},$ where observables $a$ and $b$ are incompatible,
a contextual statistical model can be projected to the complex Hilbert space.
Our model can be completed by the realist interpretation of reference observables $a\in {\cal O}.$
By the V\"axj\"o interpretation reference observables are interpreted as {\it properties of contexts:}

``If an observation of $a$ under a complex of physical
conditions $C \in {\cal C}$ gives the result $a=\alpha,$  then this value is interpreted as
the objective property of the context $C$ (at the moment of the observation).''
\subsection{Systems, ensemble representation}
 We now complete the
contextual statistical model
by considering systems $\omega$ (e.g., physical or cognitive, or social,..), cf. Ballentine (2001).
In our approach systems as well as contexts are considered as {\it elements of realty. }
In our model a context $C \in {\cal C}$ is represented  by an ensemble $S_C$ of systems which have
been interacted  with $C.$ For such systems we shall use notation:
$
\omega \hookleftarrow C
$
The set of all (e.g., physical or cognitive, or social)
systems which are used to represent all contexts $C\in {\cal C}$ is denoted by the symbol
$\Omega\equiv \Omega({\cal C}).$
Thus we have a map:
\begin{equation}
\label{VMM}
C \to S_C=\{ \omega\in \Omega:  \omega \hookleftarrow C \}.
\end{equation}
This is the ensemble representation of contexts. We set
$$
{\cal S}\equiv {\cal S}({\cal C})=\{S: S=S_C, C \in {\cal C}\}.
$$
This is the collection of all ensembles representing contexts belonging to ${\cal C}.$
The ensemble representation of contexts is given by the map (\ref{VMM})
$$
I: {\cal C} \to {\cal S}
$$
Reference observables ${\cal O}$ are now interpreted as observables on systems $\omega\in \Omega.$
In our approach it is not forbidden to interpret the values of the {\it reference observables} as objective properties
of systems.\footnote{These objective properties coexist in nature and they can be related to individual systems $\omega \in \Omega.$ However, the probabilistic description is possible only with respect to a fixed context $C.$ Noncontextual probabilities have no meaning. So values $a(\omega)$ and $b(\omega)$ coexist for a single system $\omega\in \Omega,$
but in general noncontextual (``absolute'')  probabilities ${\bf P}(\omega \in \Omega: a(\omega)=y), ...$ are not defined.
Thus, instead of mutual exclusivity of observables (cf. Bohr's principle of complementarity), we consider contextuality of probabilities and ``supplementarity'' of the reference observables (in the sense that they give supplementary statistical
information about contexts). In particular, we can speak about ``supplementarity of minds.'' Such a supplementarity
does not imply mutual exclusivity of minds; they just complete each other. It might be better to change terminology and speak about supplementarity of mental reference observables and not incompatibility.}

{\bf Definition 2.2.} {\it The ensemble representation of a contextual  statistical model
$M =({\cal C}, {\cal O}, {\cal D}({\cal O}, {\cal C}))$ is a triple
 \begin{equation}
 \label{VM1}
 S(M) =({\cal S}, {\cal O}, {\cal D}({\cal O}, {\cal C}))
 \end{equation}
 where ${\cal S}$ is a set of ensembles of systems representing contexts ${\cal C}$,
 ${\cal O}$ is a  set of observables, and ${\cal D}({\cal O}, {\cal C})$ is probabilistic data
 about ensembles ${\cal S}$ obtained with the aid of observables ${\cal O}.$}

 \subsection{Applications of V\"axj\"o model in cognitive science, psychology, sociology}
 Our contextualist
 statistical realistic models can be used not only in physics, but in any domain
 of natural and social sciences. Besides of complexes of physical conditions, we can consider
 complexes of biological, cognitive, social, economic,... conditions -- contexts -- as elements of reality.
 Such elements of reality are represented by probabilistic data obtained with the aid of
 reference observables (biological, mental, social, economic,...).

 In the same way as in physics in some special cases
 it is possible to encode such data by complex amplitudes. In this way we obtain
 representations of some biological, cognitive, social, economic,.... models in complex Hilbert spaces.
 We call them {\it complex quantum-like models.} These models describe the usual $\cos$-interference
 of probabilities. We recall again that such a representation is based on a generalized
 formula of total probability having the interference term, see section 3.

 \section{Test of quantum-like structure of mental statistics}

\subsection{Cognitive and social contexts}

 We consider examples of  cognitive contexts:

 1). $C$ can be some selection procedure which is used to select a special group $S_C$ of people or animals.
 Such a context is represented by this group $S_C$ (so this is an ensemble of cognitive systems).  For example, we select a group $S_{\rm{prof.math.}}$
 of professors of mathematics
 (and then ask questions $a$ or (and) $b$ or give corresponding tasks).  We can select a group of
 people of some age. We can select a group of people having a ``special mental state'':
 for example, people in love  or hungry people (and then ask questions or give tasks).

 2). $C$ can be a learning procedure which is used to create some special group of people or animals.
 For example, rats can be trained to react to  special stimulus.

 3). $C$ can be a collection of painting, $C_{\rm{painting}},$ (e.g. the collection of Hermitage in Sankt-Peterburg) and people interact with $C_{\rm{painting}}$
 by looking at pictures(and then there are asked questions about this collection to those people).

 4). $C$ can be, for example, ``context of classical music", $C_{\rm{cl.mus.}},$
 and people interact with $C_{\rm{cl.mus.}}$ be listening in to this music.
 In principle, we need not use an  ensemble of different people. It can be one person whom we ask questions each time
 after he has listened in to CD (or radio) with classical music. In the latter case we should use not ensemble,
 but frequency (von Mises) definition of probability.

 The last  example is an important illustration why from the beginning
 we prefer to start with the general contextualist ideology and only then we consider the possibility
 to represent  contexts by ensembles of systems.
 A cognitive context should not be identified with an ensemble of cognitive systems representing this context.
 For us $C_{\rm{cl.mus.}}$ is by itself an  element of reality.

 We can also consider {\it social contexts.} For example, social classes: proletariat-context, bourgeois-context;
 or war-context, revolution-context, context of economic depression, poverty-context
 and so on. Thus our model can be used
  in social and political sciences (and even in history). We can try to find quantum-like statistical
  data in these sciences.

\subsection{Observables}
We describe mental interference experiment.

 Let $a=x_1, x_2$ and $b=y_1, y_2$ be two dichotomous mental observables:
 $x_1$=`yes', $x_2$=`no', $y_1$=`yes', $y_2$=`no'.
 We set $X\equiv X_a= \{x_1, x_2\}, Y\equiv X_b= \{y_1,y_2\}$ (``spectra'' of observables $a$ and $b).$
 Observables can be two  different questions or two different types of cognitive tasks.
 We use these two fixed reference observables for probabilistic representation of cognitive contextual reality given by $C.$
 \footnote{Of course, by choosing another set of reference observables in general we shall obtain another representation
 of cognitive contextual reality. Can we find two fundamental mental observables? It is a very hard
 question. In physics everything is clear: the position and momentum give us the fundamental
 pair of reference observables. Which mental observables can be chosen as  mental analogous of
 the position and momentum?  In some approaches to the quantum mechanics (e.g. in Bohmian mechanics)
 position is considered as a fundamental observable, see De Broglie (1964) and D. Bohm (1951); momentum is defined as the conjugate variable. Thus we are looking for a mental analog of position, mental position.}

 \subsection{ Quantum-like structure of experimental mental data}
We perform observations of $a$ under the complex of cognitive conditions $C:$
 $$
 p^a(x)= \frac{\mbox{the number of results}\; a=x}{\mbox{the total number of observations}}, \;\; x\in X .
 $$
 So $p^a(x)$ is the probability to get the result $x$ for observation of the $a$
 under the complex of cognitive conditions $C.$
 In the same way we find probabilities $p^b(y)$ for the $b$-observation  under the same cognitive context $C.$
 \footnote{Probabilities can be ensemble probabilities or they can be time averages for measurements over one concrete
 person (e.g., each time after listening in to classical music). Measurements can be even {\it self-measurements.}
 For example, I can ask myself  questions $a$ or $b$ each time when I fall in love. These should be ``hard questions''
 (incompatible questions). By giving, e.g., the answer $a= `yes',$ I should make some important decision.
 It will play an important role when I shall  answer to the
 subsequent question $b$ and vice versa.}

 As was supposed in section 2.1 (Axiom 1), there can be created cognitive contexts $C_y$ corresponding to selections with respect
 to fixed values of the $b$-observable. The context $C_y$ (for fixed $y \in Y)$ can be characterized in the following way. By measuring the $b$-observable under the cognitive context $C_y$
 we shall obtain the answer $b= y$ with probability one.
 We perform now the $a$-measurements under cognitive contexts $C_y$ for $y= y_1, y_2,$
 and find the probabilities:
 \[p^{a/b}(x/y)=\frac{\mbox{the number of the result} \; a=x \;\mbox{under context} \;  C_y}{\mbox{the
 total number of observations} \;\mbox{under context} \;  C_y}, \;\; x \in X, y \in Y.\]

 For example, by using the ensemble approach to probability we have that the probability $p^{a/b}(x_1/y_2)$ is obtained
 as the frequency of the answer $a=x_1=`yes'$ in the ensemble of cognitive system that have already
 answered $b=y_2=`no'.$  Thus we first select a subensemble of cognitive systems who replies $`no'$ to the
 $b$-question, $C_{b=no}.$ Then we ask systems belonging to $C_{b=no}$ the $a$-question.

 It is assumed (and this is a very natural assumption) that a cognitive system is
 ``responsible for her (his) answers.'' Suppose that a system $\tau$ has answered $b=y_2=`no'.$
 If we ask $\tau$ again the same question $b$ we shall get the same answer $b=y_2=`no'.$  This is nothing else than the mental form of the von Neumann projection postulate, see von Neumann (1955): the second
measurement of the same observable, performed immediately after the first one, will yield the same value of
the observable (also Dirac (1933)); see section 3.4 for details.

 The classical probability theory tells us that all these probabilities have to be connected by the so
 called {\it formula of total probability,} see, e.g. Shiryayev (1991):
 \[p^a(x)= p^b(y_1)p^{a/b}(x/y_1) + p^b(y_2) p^{a/b}(x/y_2), \;\; x \in X .\]
 However, if the theory is quantum-like, then we should obtain  Khrennikov (2001, 2002, 2003) the  formula of total probability with an
 interference term:
 \begin{equation}
 \label{LLL}
 p^a(x)= p^b(y_1)p^{a/b}(x/y_1) + p^b(y_2) p^{a/b}(x/y_2)
 \end{equation}
 \[+ 2  \lambda(a=x/b, C) \sqrt{p^b(y_1)p^{a/b}(x/y_1)p^b(y_2) p^{a/b}(x/y_2) },\]
 where the coefficient of incompatibility (the coefficient of interference) is given by
 \begin{equation}
 \label{LLLT}
 \lambda(a=x/b, C)=\frac{ p^a(x)- p^b(y_1)p^{a/b}(x/y_1) - p^b(y_2) p^{a/b}(x/y_2) }
  {2\sqrt{p^b(y_1)p^{a/b}(x/y_1)p^b(y_2) p^{a/b}(x/y_2)}}
\end{equation}
This formula holds true for {\it incompatible observables.} To prove its validity, it is sufficient to put the expression for $\lambda(a=x/b, C),$ see (\ref{LLLT}), into (\ref{LLL}).

 In the  quantum-like statistical test for a cognitive context $C$ we
 calculate
 $$
  \lambda(a=x/b, C)=\frac{ p^a(x)- p^b(y_1)p^{a/b}(x/y_1) - p^b(y_2) p^{a/b}(x/y_2) }
  {2\sqrt{p^b(y_1)p^{a/b}(x/y_1)p^b(y_2) p^{a/b}(x/y_2)}}.
 $$
An empirical situation with $ \lambda(a=x/b, C) \not =0$ would yield evidence for quantum-like
behaviour of cognitive systems. In this case, starting with (experimentally calculated) coefficient of interference $\lambda(a=x/b, C)$
we can proceed either to the conventional Hilbert space formalism (if this coefficient is bounded by 1)
or to so called hyperbolic Hilbert space formalism (if this coefficient is larger than  1).
In the first case the coefficient of interference can be represented in the trigonometric form
 $$
 \lambda(a=x/b, C)= \cos \theta(x),
 $$
 Here $\theta(x)\equiv \theta(a=x/b, C)$ is the phase of the $a$-interference between
 cognitive contexts $C$ and $C_y, y \in Y.$ In this case we have the conventional formula of total probability with the interference term:
  \begin{equation}
 \label{LLLQ}
 p^a(x)= p^b(y_1)p^{a/b}(x/y_1) + p^b(y_2) p^{a/b}(x/y_2)
 \end{equation}
 \[+ 2  \cos \theta(x) \sqrt{p^b(y_1)p^{a/b}(x/y_1)p^b(y_2) p^{a/b}(x/y_2) }.\]
 In principle, it could be derived in the conventional Hilbert space formalism. But we chosen the inverse way.
Starting with (\ref{LLLQ}) we could introduce a ``mental wave function''
 $\psi\equiv \psi_C$ (or pure quantum-like mental state) belonging to this Hilbert space, see section 5. We recall that in our approach
 a `mental wave function' $\psi$  describes cognitive context $C.$ This is nothing else than a special mathematical
 encoding of probabilistic information about this context which can be obtained with the aid of reference observables
 $a$ and $b.$

\subsection{Von Neumann postulate in cognitive science and psychology}
For further considerations (on a wave function) it is important to underline that in general all above quantities can depend on a cognitive context  $C:$
$$
 p^a(x) =p_{C}^a(x), p^b(y) =p_{C}^b(y),
 p^{a/b}(x/y)=p^{a/b}_{C}(x/y), \theta(x)=\theta_{C}^{a/b}(x).
$$
Dependence of probabilities $p_{C}^a(x), p_{C}^b(y)$ on a context $C$ is irreducible (these probabilities are defined by
$C).$ But in some cases dependence of the transition probabilities $p^{a/b}_{C}(x/y)$ on $C$ could be reducible.
In the experimental situation these probabilities (frequencies) are found in the following way. First cognitive systems
interact with a context $C.$ In this way there is created an ensemble $S_C$ of cognitive systems representing the context $C.$ Then cognitive systems belonging to the ensemble $S_C$ interact with a new context $C_y$  which is determined by the mental observable $b.$ For example, students belonging to a group $S_C$ (which was trained under the mental or social conditions $C)$  should answer to the question $b.$ If this question is so disturbing for a student
$\omega$ that he would totally forget about the previous $C$-training, then the transition probabilities do not depend on $C: p^{a/b}(x/y).$

We remark that this is the case in conventional quantum theory. Here for incompatible (noncomutative) observables the transition probabilities $p^{a/b}(x/y)$ do not depend on the previous context $C,$ i.e.,  a context preceding the $b=y$ filtration. In quantum theory any $b=y$ filtration destroys the memory on the preceding
physical context $C.$  This is our contextual interpretation of the von Neumann projection postulate.

We do not know the general situation for cognitive systems.\footnote{ It might be that the von Neumann projection postulate
can be violated by cognitive systems. In such a case we would not be able to construct the conventional quantum
representation of contexts by complex probability amplitudes, cf. section 5.}  Our conjecture is that:

\medskip

{\bf Postulate.} (``von Neumann postulate for mental observable'') {\it For any pair $a, b$ of incompatible mental observables the transition probability $p^{a/b}(x/y)$ is completely determined by the preceding preparation -- context $C_y$ corresponding to the $[b=y]$-filtration.}

\medskip

In principle, we could be satisfyed even by a weaker form of this postulate.

\medskip

{\bf Postulate.} (``Weak von Neumann postulate for mental observable'') {\it There exist incompatible mental observables
$a, b$ such that the transition probability $p^{a/b}(x/y)$ is completely determined by the preceding preparation -- context $C_y$
corresponding to the $[b=y]$-filtration.}

Finally, we remark that in our contextual approach the von Neumann postulate (for physical as well as mental systems)
is not so mysterious. This is nothing else as the condition of Markovness for successive measurements.

\section{Hyperbolic interference of minds}
As was already mentioned, statistical
data obtained in experiments with cognitive systems could produce
the coefficient of interference which is larger than 1.
In general  quantities
$$
\lambda(a=x/b, C)=
\frac{ p^a(x)- p^b(y_1)p^{a/b}(x/y_1) - p^b(y_2) p^{a/b}(x/y_2) }
  {2\sqrt{p^b(y_1)p^{a/b}(x/y_1)p^b(y_2) p^{a/b}(x/y_2)}}.
$$
can extend 1 (see Khrennikov (2001, 2002, 2003) for examples). In this case we can introduce a hyperbolic phase
parameter $\theta\in [0, \infty)$ such that
$$
\cosh \theta(x)=\pm \frac{ p^a(x)- p^b(y_1)p^{a/b}(x/y_1) - p^b(y_2) p^{a/b}(x/y_2) }
  {2\sqrt{p^b(y_1)p^{a/b}(x/y_1)p^b(y_2) p^{a/b}(x/y_2)}}.
$$
In this case we can not proceed to the ordinary Hilbert space formalism. Nevertheless, we can use an
analog of the complex Hilbert space representation for probabilities.
Probabilities corresponding such cognitive contexts
can be represented in a hyperbolic Hilbert space -- module over a two-dimensional Clifford algebra, see
Khrennikov (2001, 2002, 2003).\footnote{ At the moment there are no experimental confirmations of hyperbolic
interference for cognitive systems. If such a result was obtained it would imply that cognitive systems
have more rich probabilistic structure than quantum systems.}

In principle it may occur that $\vert \lambda_1\vert\leq 1$ and $\vert \lambda_2\vert > 1$
or vice versa. In this case we obtain {\it hyper-trigonometric interference} of minds.

\section{Mental wave function}
Let $C$ be a cognitive context.
We consider only cognitive contexts  with trigonometric interference for {\it incompatible mental observables}
$a$ and $b.$ It is assumed that the Weak von Neumann postulate for mental observable holds for $a$ and $b.$
The interference formula of total probability (\ref{LLL}) can be written in
the following form:
\begin{equation}
\label{Two}
p_{C}^a(x)=\sum_{y \in Y} p_{C}^b(y) p^{a/b}(x/y) +
2\cos \theta_{C}(x)\sqrt{\Pi_{y \in  Y}p_{C}^b(y) p^{a/b}(x/y)}
\end{equation}
By using the elementary formula:
$$
D=A+B+2\sqrt{AB}\cos \theta=\vert \sqrt{A}+e^{i\theta}\sqrt{B}|^2, A, B>0,
$$
we can represent the probability $p_C^b(x)$ as the square of the complex amplitude:
\begin{equation}
\label{Born}
p_{C}^a(x) =\vert\psi_{C}(x)\vert^2
\end{equation}
where
\begin{equation}
\label{EX}
\psi(x)\equiv \psi_{C}(x)=\sum_{y\in Y}\sqrt{p_{C}^b(y)p^{a/b}(x/y)}
e^{i \xi_{C}(x/y)} .
\end{equation}
Here phases $\xi_{C}(x/y)$ are such that
\[\xi_{C}(x/y_1) - \xi_{C} (x/y_2)=\theta_{C}(x) .\]
We denote the space of functions: $\psi: X \to {\bf C}$ by the symbol
$E=\Phi(X, {\bf C}).$ Since $X= \{x_1, x_2 \},$ the $E$ is the two dimensional
complex linear space. Dirac's $\delta-$functions $\{ \delta(x_1-x), \delta(x_2-x)\}$
form the canonical basis in this space. For each $\psi \in E$ we have
\[\psi(x)=\psi(x_1) \delta(x_1-x) + \psi(x_2) \delta(x_2-x).\]

Denote by the symbol ${\cal C}^{\rm{tr}}$ the set of all cognitive contexts
having the trigonometric statistical behaviour (i.e., $\vert \lambda \vert \leq 1)$
with respect to mental observables $a$
and $b.$ By using the representation (\ref{EX}) we construct the map
$$
\tilde{J}^{a/b}:{\cal C}^{\rm{tr}} \to \tilde{\Phi}(X, {\bf C}),
$$
where $\tilde{\Phi}(X, {\bf C})$ is the space of equivalent classes of
functions under the equivalence relation: $\varphi$ equivalent   $\psi$ iff
$\varphi= t\psi, t \in {\bf C,} \vert t\vert=1.$

To fix some concrete representation of a context $C,$
we can choose, e.g., $\xi_{C}(x/y_1)=0$ and $\xi_{C}(x/y_2)
=\theta_{C}(x).$ Thus we construct the map
\begin{equation}
\label{R}
J^{a/b}:{\cal C}^{\rm{tr}}\to \Phi(X, {\bf C})
\end{equation}
The $J^{a/b}$ maps cognitive contexts
into complex amplitudes. The representation ({\ref{Born}}) of probability as the square of the
absolute value of the complex $(a/b)-$amplitude is nothing other than the
famous {\bf Born rule.} The complex amplitude $\psi_{C}$ can be called a {\it mental wave function}
or pure mental state (QL-state).

We emphasize that the map $J^{a/b}$ is not surjective. It can happen
that
$J^{a/b}(C_1)= J^{a/b}(C_2)$ for different context $C_1$ and $C_2$
(if $p_{C_1}^a(x)= p_{C_2}^a(x)$ and
$p_{C_1}^b(y)= p_{C_2}^b(y)).$ Such contexts are represented by the same complex amplitude
$\psi(x)= \psi_{C_1}(x)= \psi_{C_2}(x).$ In particular, it can be that different cognitive contexts
are represented by the same mental wave function $\psi.$ We shall come back to this problem  in section 6. We set
\[e_x^a(\cdot)=\delta(x- \cdot)\]
The representation (\ref{Born}) can be rewritten in the following form:
\begin{equation}
\label{BH}
p_{C}^a(x)=\vert(\psi_{C}, e_x^a)\vert^2 \;,
\end{equation}
where the scalar product in the space $E=\Phi(X, C)$ is defined by the
standard formula:
\[(\varphi, \psi)=\sum_{x\in X} \varphi(x)\bar \psi(x) .\]
The system of functions $\{e_x^a\}_{x\in X}$ is an orthonormal basis in the
Hilbert space $H=(E, (\cdot, \cdot))$

Let $X \subset {\bf R},$ where ${\bf R}$ is the set of real numbers. By using the Hilbert space representation of
Born's rule ({\ref{BH}}) we obtain for the Hilbert space representation of the classical conditional
expectation:
\begin{equation}
\label{BI1}
E (a/C) = \sum_{x\in X} x p^a(x)=\sum_{x\in X} x \vert \psi_{C}(x)\vert^2
=(\hat a \psi_{C},  \psi_{C}) \;,
\end{equation}
where $\hat a:\Phi(X,{\bf C})\to \Phi(X, {\bf C})$
is the multiplication operator. This operator can also be determined by its
eigenvectors: $\hat a e_x^a=x e_x^a, x\in X.$

We notice that if the matrix of transition probabilities $P^{a/b}=(p^{a/b}(x/y))$ is {\it double
stochastic} we can represent the mental observable $b$ by a symmetric operator $\hat b$ in the
same Hilbert space.  In general operators $\hat a$ and $\hat b$ do not commute, Khrennikov (2003).

\section{Quantum-like projection of mental reality}
We emphasize that the quantum-like representation is created through a projection of underlying mental realistic model
to the complex Hilbert space. Such a projection induces a huge loss of information about the underlying mental model.
Thus the quantum-like model gives a very rough image of the realistic  model.
\subsection{Social opinion pull}
Let us consider a  family of social contexts ${\cal C}$ such that each context correspond to society of some country:
$C_{\rm{USA}}, C_{\rm{GB}}, C_{\rm{FR}}, ..., C_{\rm{GER}},...$ and let us consider two reference observables given by questions:

\medskip

a). ``Are you against pollution?''

b). ``Would you want to have lower prices for gasoline?''

\medskip

It is supposed that observables $a$ and $b$ are incompatible:
$$
p^{a/b} (a= \rm{yes}/ b= \rm{no})\not=0,
p^{a/b} (a= \rm{no}/ b= \rm{no})\not=0,
$$
$$
p^{a/b} (a= \rm{yes}/ b= \rm{yes})\not= 0, p^{a/b} (a= \rm{no}/ b= \rm{yes})\not= 0.
$$
Thus the corresponding social groups in the society are assumed to be statistically essential.
Moreover, to get a quantum-like representation it should be assumed that the transition probabilities
$p^{a/b} (a= x/ b= y)$ do not depend on a society $C.$ For example, the proportion of people who are
against pollution among people who are satisfied by prices for gasoline is the same in USA, Great Britain,
France and so on. Of course, this is a rather strong assumption. And its validity should be checked.

The pull under consideration contains very important questions. In our  QL-model in those
societies are represented by complex probability amplitudes
$$
\psi_{\rm{USA}}, \psi_{\rm{GB}}, \psi_{\rm{FR}}, ..., \psi_{\rm{GER}},...
$$
These mental wave functions can be used to investigate some essential features of these societies. But, of course,
answers to the questions $a$ and $b$ do not completely characterize a society. Thus the quantum-like representation induces
the huge loss of information.
\subsection{Quantum-like representation of functioning of neuronal structures}
Let us consider two coupled neural networks $N_1$ and $N_2.$ We assume that they are
strictly hierarchic in the sense that there are ``grandmother'' neurons $n_1$ and $n_2$
in networks $N_1$ and $N_2,$ respectively.\footnote{The model of cognition based on
grandmother neurons was dominating in 60 -- 80th, Amit (1989) . Later it was strongly criticized,
but was not totally rejected. In the modified approach there are considered grandmother neuronal groups,
instead of single neurons.} The integral network $N=N_1+N_2$ interacts with  contexts
$C$ which are given by input signals into both networks. For example, contexts ${\cal C}=\{ C\}$ can be  visual images and the integral network $N$ recognizes  those images (e.g. $N_1$ is responsible for countors and $N_2$  for
colors). We use so called frequency-domain approach, see for example Hoppenstead (1997), and assume that cognitive information is presented
by frequencies of firing of neurons. Consider two reference observables $a, b$ where $a=1: n_1 $ firing, and  $a=0: n_1 -$nonfiring,  and $b=1: n_2-$firing, and  $b=0: n_2-$nonfiring. Our quantum-like formalism gives the possibility to represent each context $C$ (e.g., an image $C)$ by a complex probability amplitude $\psi_C.$ Here probabilities
${\bf P}(a=x/C), {\bf P}(b=y/C)$ are defined as frequencies. Such an amplitude can be reconstructed on the basis
of measurements on grandmother neurons $n_1$ and $n_2.$ Of course, $\psi_C$ gives only a projection of the neuronal image
of the context $C.$  The complete neuronal image is given by frequencies of firing of all neurons in the network
$N$ and the QL-image $\psi_C$ is based only on frequencies of firings of grandmother neurons.  However, we could not exclude that cognition (and consciousness) is really based on such a QL-projecting of neronal
states, see section 7.
\subsection{Quantum-like representation of Freud's psychoanalysis}
In its original form Freud's psychoanalysis was based on the representation of psychical states of patients through
two groups of questions:

\medskip

$a:$ about recollections from childhood;

$b:$ about sexual experiences.

\medskip

Later Sigmund Freud was  criticized for his attempt to reduce the psychical state to the $a,b$-domains.
He was especially strongly critisized for overestimating the role of sexual experiences in childhood.
From our point of view Freud's approach might be considered as the basis for creation of a
quantum-like representation based on the reference observables $a$ and $b.$ Of course, a mental
$\psi$-function based on observables $a$ and $b$ gives a very rough representation of the underlying
mental context of a patient, but nevetheless it could be used to proceed mathematically.

\section{Quantum-like consiousness}
The brain is a huge information systems which contains millions of minds. It could not ``recognize''
(or ``feel'') all those minds  at each instant of time $t.$\footnote{It may be more natural to consider mental (or psychological) time and not physical time, see e.g. Khrennikov (2000). There are experimental evidences that: a) cognition is not based on the continuous time processes (a moment in mental time correlates with $\Delta\approx 100
ms $ of physical time);
b) different psychological functions operate on different scales of physical time. In Krennikov (2000) mental time
was described mathematically by using $p$-adic hierarchic trees.} Our fundamental hypothesis is that the brain is able
to create the QL-representations of minds. At each instant of time $t$ the brain creates the  QL-representation of its mental context $C$ based on two incompatible\footnote{As was mentioned in footnote 14, see section 2.2, it is more natural to call $a$ and $b$ supplementary and incompatible.} mental (self-)observables $a$ and $b.$ Here
$a=(a_1,..., a_n)$ and $b=(b_1,..., b_n)$ can be very long vectors of compatible dichotomous observables. The
(self-)reference observables can be chosen (by the brain) in different ways at different instances of time. Such a
change of the reference observables is known in cognitive sciences as a {\it change of representation.}

A mental context $C$ in the $a/b-$ representation is described by the mental wave function $\psi_C.$ We can speculate
that the brain has the ability to feel this mental field as a distribution
on the space $X.$ This distribution is given by the norm-squared of the mental wave function: $\vert \psi_C(x) \vert^2.$
This mental QL-wave  contributes into the deterministic dynamics of minds, e.g. by inducing Bohmian
quantum potential, see e.g. Khrennikov (2000).

In such a model it might be supposed that the state of our consciousness is represented by the mental wave function
$\psi_C.$ By using Freud's terminology we can say that one has classical {\it subconsiousness} and quantum-like consiousness, cf. Khrennikov (2000). QL-consiousness is represented by the mental wave function $\psi_C.$ The crucial point
is that in this model consciousness is created through neglecting an essential volume of information contained in subconsciousness. Of course, this is not just a random loss of information. Information is selected through the algorithm presented in section 5: context $C$ is projected onto $\psi_C.$

The (classical) mental state of subconsiousness evolves with time $C\to C(t).$ This dynamics induces dynamics of
the mental wave function $\psi(t)=\psi_{C(t)}$ in the complex Hilbert space, see Khrennikov (2004) for the mathematical details.

\medskip

{\bf Postulate QLR.} {\it The brain is able to create the QL-representation of mental contexts, $C\to \psi_C$
(by using the algorithm based on the formula of total probability with interference, see section 5).}

\section{Brain as quantum-like computer}
We can speculate that the ability of the brain to create the QL-representation of mental contexts, see
Postulate QLR, induces functioning of the brain as a quantum-like computer.

\medskip

{\bf Postulate QLC.} {\it The brain performs computation-thinking by using algorithms of quantum computing
in the complex Hilbert space of mental QL-states.}

\medskip

We emphasize that in our approach the brain is not  quantum computer, but QL-computer. On one hand, QL-computer
works totally in accordance with mathematical theory of quantum computations (so by using quantum algorithms). On the other hand, it is not based on superposition of individual mental states. The complex amplitude $\psi_C$ representing a mental context $C$ is a special probabilistic representation of information states of the huge neuronal ensemble.
In particular, the brain is {\it macroscopic} QL-computer. Thus the QL-parallelism (in the opposite to conventional quantum parallelism) has a natural realistic base. This is real parallelism in working of millions of neurons. The crucial point is the way in which this classical parallelism is projected onto dynamics of QL-states. The QL-brain is able to solve  NP-problems. But there is nothing mysterious in this ability: exponentially increasing number of operations is
performed through involving of exponentially increasing number of neurons.

We pay attention that by coupling QL-parallelism to working of neurons we started to present a particular ontic model
for QL-computations. We shall discuss it in more detail. Observables $a$ and $b$ are self-observations of brain. They can be represented as functions of the internal state of brain $\omega.$ Here $\omega$ is a parameter of huge dimension
describing states of all neurons in brain: $\omega= (\omega_1, \omega_2,..., \omega_N):$
$$
a=a(\omega), b=b(\omega).
$$
The brain is not interested in concrete values of the reference observables at fixed instances of time. The brain
finds the contextual probability distributions $p_C^a(x)$ and $p_C^b(y)$ and creates the mental QL-state $\psi_C(x),$
see algorithm in section 5.
Then it works with $\psi_C(x)$ by using algorithms of quantum computing. The crucial problem is to find  mechanism
of calculating of contextual probabilities. We think that they are frequency probabilities which are created in the brain in the following way.

There are two scales of time: a) internal scale; b) QL-scale. The internal scale is finer than the QL-scale.
Each instant of QL-time $t$ corresponds to an interval $\Delta$ of internal time $\tau.$ We might identify the QL-time with mental (psychological) time and the internal time with physical time. During the interval $\Delta$ of internal time the brain collects statistical data for self-observations of $a$ and $b.$

Thus the internal state $\omega$ of the brain evolves as $\omega= \omega(\tau, \omega_0).$ At each instance of internal time $\tau$ there are performed nondisturbative self-measurements of $a$ and $b.$ These are realistic measurements: the brain gets values $a(\omega(\tau, \omega_0)),$ $b(\omega(\tau, \omega_0)).$ By finding frequencies of realization of fixed values for $a(\omega(\tau, \omega_0))$ and  $b(\omega(\tau, \omega_0))$ the brain obtains the frequency probabilities
$p_C^a(x)$ and $p_C^b(y).$ These probabilities are related to the instant of QL-time time $t$ corresponding to the interval of internal time $\Delta: p_C^a(t, x)$ and $p_C^b(t,y).$

For example, $a$ and $b$ can be measurements over different  domains of brain. It is supposed that the brain can ``feel'' probabilities
(frequencies) $p_C^a(x)$ and $p_C^b(y),$ but not able to ``feel'' the simultaneous probability distribution
$p_C(x,y) = P(a=x, b=y/C).$ This is not the problem of mathematical existence of such a distribution.\footnote{We recall
 that, since we consider only two realistic observables, there is no direct contradiction with Bell's inequality.}
 This is the problem of integration of statistics of observations from different domains of the brain.
By using the QL-representation based only on probabilities $p_C^a(x)$ and $p_C^b(y)$  the brain could be able to
escape integration of information about {\it individual self-observations} of variables $a$ and $b$
related to spatially separated domains of brain. The brain  need not couple these domains at each instant of internal time $\tau.$ It couples them only once in the interval $\Delta$ through the contextual probabilities
$p_C^a(x)$ and $p_C^b(y).$ This induces the huge saving of time.

\section{Evolution of mental wave function}
The mental wave function $\psi(t)$ evolves in the complex Hilbert space (space of probability amplitudes, see section 5). The straightforward generalization of quantum mechanics would imply the {\it linear} Schr\"odinger equation:
\begin{equation}
\label{LUB}
i\frac{d\psi(t)}{d t} = \hat{\cal H} \psi(t), \psi(0)= \psi_0,
\end{equation}where $ \hat{\cal H} : H \to H$ is a self-adjoint operator in the Hilbert space $H$
 of mental QL-states. However, the V\"axj\"o model predicts, Khrennikov (2004),  broader spectrum of evolutions
 in the Hilbert space (induced by evolutions of contexts). We could not go deeply into mathematical details and only remark that in general the contextual dynamics $C\to C(t)$ can induce {\it nonlinear} evolutions in $H:$
\begin{equation}
\label{LUB1}
i\frac{d\psi(t)}{d t} = \hat{\cal H}(\psi(t)), \psi(0)= \psi_0,
\end{equation}
where $ \hat{\cal H} : H \to H$ is a nonlinear map. It is important to point out that even the nonlinear dynamics
in the Hilbert state space induced by a contextual dynamics is {\it unitary:}
$(\psi(t), \psi(t)) = (\psi(0), \psi(0)).$

In principle,

\medskip

{\it there are no a priory reasons to assume that the mental quantum-like dynamics should always be linear!}

\medskip

It might be that nonlinearity of the Hilbert space dynamics is the distinguishing feature of cognitive systems.
 However, at the present time this is just a speculation. Therefore it would be interesting to consider a linear mental quantum-like dynamics.\footnote{In any event linear dynamics can be considered as an approximation of nonlinear dynamics.} For example, let us consider a
quantum-like Hamiltonian:
\begin{equation}
\label{LUB2}\hat{\cal H} \equiv {\cal H}(\hat a, \hat b)=\frac{{\hat b}^2}{2} + V (\hat a),
\end{equation}
where $V: X \to {\bf R}$ is a ``mental potential'' (e.g. a polynomial), cf Khrennikov (1999, 2000, 2003).
We call $\hat{\cal H}$ the operator of {\it mental energy.}
 Denote by
$\psi_j$ stationary mental QL-states: $\hat{\cal H} \psi_j= \mu_j \psi_j.$ Then any mental QL-state
$\psi$ can be represented as a superposition of stationary states:
\begin{equation}
\label{LUB3}
\psi= k_1 \psi_1 + k_2 \psi_2, \; k_j \in {\bf C}, \vert k_1\vert^2 +\vert k_2\vert^2 =1.
\end{equation}
One might speculate that the brain has the ability to feel superpositions (\ref{LUB3}) of stationary
mental QL-states. In such a case superposition would be an element of mental reality.
However, it seems not be the case. Suppose that $\psi_1$ corresponds to zero mental energy, $\mu_1=0.$
For example, such a QL-state can be interpreted as the state of depression. Let $\mu_2 >>0.$ For example,
such a QL-state can be interpreted as the state of excitement. My internal mental experience tells
that I do not have a feeling of superposition of states of depression and high excitement. If I am not
in one of those stationary states, then I am just in a new special mental QL-state $\psi$ and I have the feeling of this $\psi$ and not superposition.\footnote{We exclude abnormal behavior such as manic-depressive syndrome.} Thus it seems
that the expansion (\ref{LUB3}) is just a purely mathematical feature of the model.

\section{Noninjectivity of correspondence between classical subconsciousness and quantum-like consciousness}
We use here the interpretation proposed in the previous section and pay attention that the map $J^{a/b}:{\cal C}^{\rm{tr}}\to \Phi(X, {\bf C}),$ see section 5, is not one-to-one. Thus it can be that a few different contexts $C, C^\prime,..$ are represented by the same mental QL-state $\psi.$ Suppose now that this is a stationary state --
an eigenstate of the operator of mental energy. We pay attention that in general the corresponding mental context is not uniquely determined. QL-stationarity of a state $\psi_j$ can be based on a rather complex dynamics of context,
$C_j(t),$  in subconsiousness.

\section{Structure the set of states  of mental systems}

We recall few basic notions of the statistical formalism of quantum theory, see,
e.g., Holevo (2001). States of quantum systems  are mathematically
represented by density operators -- positive operators $\rho$ of unit trace. Pure states (wave functions)
$\psi$ are represented by vectors belonging to the unit sphere of a Hilbert space -- corresponding
density operators $\rho_\psi$ are the orthogonal projectors onto one dimensional subspaces corresponding to vectors
$\psi.$ The set ${\cal D}$ of states (density operators) is a convex set.

In the two dimensional case (corresponding to dichotomous observables -- `yes' or `no' answers)
the set ${\cal D}$ can be represented as the unit ball in the three dimensional
real space ${\bf R}^3.$ Pure states are represented as  the unit sphere, {\it Bloch sphere.}
Here the whole set  ${\cal D}$ is the convex hall of the Bloch sphere $S$, i.e., of the set of pure states.

In our mental QL-model  some contexts (producing trigonometric interference)
are represented by points in $S.$ We can also consider statistical mixtures of these pure states.
Let $S_{{\cal C}^{\rm{tr}}}= J^{a/b}({\cal C}^{\rm{tr}})$ (the image of the set of populations ${\cal C}^{\rm{tr}}).$
Then the set of mental states ${\cal D}_{{\cal C}^{\rm{tr}}}$ coincides with the convex hall of the
$S_{{\cal C}^{\rm{tr}}}.$ There are no reasons to suppose that $S_{{\cal C}^{\rm{tr}}}$ would coincide with the Bloch sphere $S.$
Thus there is no reasons to suppose that the set of mental QL-states ${\cal D}_{{\cal C}^{\rm{tr}}}$
would coincide with the set of quantum states ${\cal D}.$
It is the fundamental problem\footnote{Of course, if you would accept our quantum-like
statistical ideology.} to describe the set of pure quantum-like metal states  $S_{{\cal C}^{\rm{tr}}}$
for various classes of cognitive systems.

We might speculate that  $S_{{\cal P}}$ depends essentially on a class of cognitive system.
So $S_{{\cal P}}^{\rm{human}}$ does not equal to $S_{{\cal P}}^{\rm{leon}}.$
We can even speculate that in the process of evolution the set $S_{{\it P}}$
have been increasing and $S_{{\cal P}}^{\rm{human}}$ is the maximal set of mental states.
It might even occur that $S_{{\cal P}}^{\rm{human}}$ coincides with the Bloch sphere.

\section{Single-mind, many minds  and V\"axj\"o approaches}
As was pointed by Barret (1999), p. 211: {\small ``Just as with many-worlds theories, there are many many-minds
formulations of quantum mechanics.''} It seems that the approach of Albert and Loewer (1988) to
the many-minds theory is the most close to our approach. It was created to provide a more natural foundations of Everett's many-worlds interpretation. One could easier accept the presence of many minds than many worlds. However, we are not
so much interested in the original aim of the many-minds approach as a mental interpretation of Everett's many-worlds
quantum mechanics. We are interested on consequences of Albert--Loewer theory for cognitive science. We start with so called single-mind theory, see e.g. Barret (1999) for compact and simple presentation:
{\small ``The single-mind theory is a sort of hidden-variable theory, but instead of taking positions as always determinate, one takes mental states as always determinate. ... This explanation of the determinacy of experience requires one to adopt a theory of mind where an observer's mental state is well defined at an instant.''} It is very important for us that  {\small ``The individual minds, as on the [single-mind theory], are not quantum mechanical systems; they are never in superposition''.} Such a viewpoint to individual minds is also characteristic for our contextual realistic mental model: a single mind could not be in a superposition of states; in particular, interference is not a self-interference of a single mind. We remark that the   Albert--Loewer and V\"axj\"o
approaches differ crucially from quantum logic approach. In the latter a single mind is in a superposition of mental states and mental interference is a self-interference of mind.

Albert and Loewer (1988, 1992) also created a many-minds theory. In this theory, see Albert and Loewer (1988), p.206, 130 and Barett (1999), p.192, : {\small ``
... every sentient physical system, every observer, is associated with not a single mind but rather a continuous infinity of minds. Each mind is supposed to evolve exactly as described in the single-mind theory; there are just more of them
associated with each observer.''} This picture does not contradict to our contextual statistical realistic model
of mental reality. However, we emphasize again (see introduction for comparison with Bohmian mechanics)
that the V\"axj\"o model is just an observational  model like the orthodox Copenhagen model.\footnote{The main difference is that the V\"axj\"o model is realistic (for the reference observables).} Therefore we are not interested in what happens really in the brain (e.g., how those single minds evolve). For us it is important only the presence of ensembles of minds in that each mind has a determinate state.

From the point of view of the many-minds model we do the following thing. Denote the collection of observer's minds
(at some instant) by $C.$ In the V\"axj\"o terminology this is a mental (or cognitive) context. Consider two incompatible
mental observables $a$ and $b$ (e.g. given in the form of questions). Then we can construct a mental wave function $\psi_C$, see section 5, representing the collection of minds $C.$

Another many-minds interpretation was proposed by Lookwood (1989, 1996). In the debate  between Lookwood (1996) and Loewer (1996)
we would choose the side of Albert and Loewer. As in the theory of Albert and Loewer (1988), we assume that minds have definite states at each instant. We also assume random dynamics of minds (in subconsiousness, see section 7). Therefore minds in the V\"axj\"o model as well as in the Albert--Loewer (1988) model {\small ``have reliable memories of their own past mental states,''} p. 121.  As was pointed by Barret (1999), the latter assumption about memories implies that
Lockwood's belief {\small ``that his theory is empirically equivalent to Albert and Loewer's''} was not justified
({\small ``since minds on Lookwood's many-minds theory do not have even transcendental identities,''} p.210).
Neither our theory could be connected to Lookwood's theory. \footnote{
Here we do not discuss many-minds theory of Donald (1990, 1995, 1996), since his approach is based on the reductionist idea that
``quantumness of mind'' is a consequence of ``quantumness of brain'' as a physical system. As was pointed out in
introduction, our model could be used neither as an argument for supporting  nor rejecting quantum reductionism.}

\section{On the notion of  macroscopic quantum system}
The study of macroscopic quantum systems
is the subject of the greatest interest for foundations of quantum mechanics as well as its applications.
However, I would like to pay attention to the fact that (at least for me)
it is not clear:

{\it ``What can be called a macroscopic quantum system?''}

Of course, this question is closely related to the old question:

{\it ``What can be called a quantum system?''}

There is no common point of view to such notions as quantization,
quantum theory.  For me (in the opposition to N. Bohr) the presence
of quanta (of, e.g., energy) is not the main distinguishing feature
of quantum theory. Of course, the presence of observables (e.g.,
energy) with discrete spectra is an important feature of  quantum
theory. However, the basic quantum observables, the position and the
momentum, still have continuous ranges of values. I think that the
main point is that quantum theory is a {\it statistical theory.}
Therefore it should be characterized in statistical terms. We should
find the basic feature of quantum theory which distinguishes this
theory form classical statistical mechanics.  The {\it interference
of probabilities} is such a basic statistical feature of quantum
theory. Therefore any system (material or not) which exhibits (for
some observables) the interference of probabilities should be
considered as a quantum system (or say ``quantum-like system'').
 Thus, since human beings by replying to special pairs of questions produce, see Conte et al. (2004),
interference of probabilities, they should be considered as macroscopic quantum systems.\footnote{I presented
this viewpoint to macroscopic quantum systems in my discussions with A. Leggett (after his public lecture
in Prague connected with the Conference ``Frontiers of Quantum and Mesoscopic Thermodynamics'', Prague, July-2004,
and  during my talk at University of Illinois). Unfortunately, neither A. Leggett nor other participants
of the conference buy my idea on human being as a macroscopic quantum system.}

As I understood, for physicists a macroscopic quantum system is
a huge ensemble of microscopic quantum systems (e.g., electrons) prepared in a special state.
Human being is also a huge ensemble of microscopic quantum systems... However, the state of this
ensemble cannot be considered as quantum from the traditional point of view. Nevertheless, according to
our interference viewpoint to quantumness human being is quantum (but not because it is composed of
microscopic quantum systems).

In the connection with our discussion on the definition of a
macroscopic quantum system it is natural to mention experiments of
A. Zeilinger and his collaborators on interference of probabilities
for fullerens and other macromolecules including bimolecular
porphyrin. It seems that A. Zeilinger uses the same definition of
quantumness as I. It is interesting that one of the main aims of
further experiments of A. Zeilinger is to find the interference of
probabilities for some viruses. I think that at that point he will
come really very close to my viewpoint to macroscopic quantum
systems, in particular, biological quantum systems.

 \bigskip

 {\bf Acknowledgements:}

I would like to thank S. Albeverio, H. Atmanspacher, E. Conte, C. Fuchs, A. Grib, A. Plotnitsky, B. Hiley, A. Holevo, G. Vitiello, G. S. Voronkov for fruitful discussions.

\bigskip

{\bf References}

Albert, D. Z., Loewer, B., 1988. Interpreting the many worlds interpretation. Synthese 77, 195-213.

Albert, D. Z., 1992. Quantum mechanics and experience. Cambridge, Mass.: Harvard Univ. Press.

Amit, D., 1989. Modeling Brain Function. Cambridge
Univ. Press, Cambridge.

Ashby, R., 1952,  Design of a brain. Chapman-Hall, London.

Atmanspacher, H., Biskop,  R. C., Amann,  A., 2001. Extrinsic and intrinsic irreversibility in
probabilistic dynamical laws. Proc. Conf.  Foundations of Probability and Physics, ed. A. Yu. Khrennikov.
Q. Prob. White Noise Anal. 13, 50-70, WSP, Singapore.

Atmanspacher, H., Kronz, F., 1999. Relative onticity. In  On quanta, mind, and matter. Hans Primas in context.
ed. H. Atmanspacher, A. Amann, U. M\"uller-Herold, 273-294, Kluwer, Dordrecht.

Ballentine, L. E., 2001. Interpretations of probability and quantum theory.
Proc. Conf. Foundations of Probability and Physics, ed. A. Yu. Khrennikov.
Q. Prob. White Noise Anal. 13, 71-84, WSP, Singapore.

Barrett,  J. A., 1999.  The quantum mechanics of minds and worlds. Oxford Univ. Press, Oxford.

Bechtel, W., Abrahamsen,  A.,  1991. Connectionism and the mind.
Basil Blackwell, Cambridge.

Bohm, D., 1951. Quantum theory. Prentice-Hall,
Englewood Cliffs, New-Jersey.

Bohm, D., Hiley,   B., 1993.  The undivided universe:
an ontological interpretation of quantum mechanics.
Routledge and Kegan Paul,  London.

Choustova, O., 2004. Bohmian mechanics for financial processes.
J. Modern Optics 51, n. 6/7, 1111.

Conte, E., Todarello, O., Federici,  A.,  Vitiello, T., Lopane,  M., Khrennikov,  A. Yu., 2004.
A Preliminar Evidence of Quantum Like Behavior in Measurements of Mental States. In
Quantum Theory: Reconsideration
of Foundations, edited by A. Yu. Khrennikov,  Ser. Math. Modeling 10,  679-702,
V\"axj\"o Univ. Press, V\"axj\"o; http://xxx.lanl.gov/abs/quant-ph/0307201.

De Broglie, L., 1964. The current interpretation of wave mechanics,
critical study. Elsevier Publ., Amsterdam-London-New York.

Dirac, P. A. M., 1930. The Principles of Quantum Mechanics.
Oxford Univ. Press, Oxford.

Donald, M. J., 1996. On many-minds interpretation of quantum mechanics. Preprint.

Donald, M. J., 1990. Quantum theory and brain. Proc. Royal Soc. A 427, 43-93.

Donald, M. J., 1995. A mathematical characterization of the physical structure of observers.
Found. Physics 25/4, 529-571.

Deutsch, D., 1997. The Fabric of Reality. How much can our four deepest theories of the world explain?
Publisher Allen Lane, The Penguin Press.

Eliasmith, C., 1996.  The third contender: a critical examination of the dynamicist
theory of cognition.  Phil. Psychology 9(4), 441-463.

Fuchs, C., 2002. Quantum mechanics as quantum information (and only a little more).
Proc. Conf. Quantum Theory: Reconsideration
of Foundations, ed. A. Yu. Khrennikov.
Ser. Math. Modelling  2, 463-543, V\"axj\"o Univ. Press,  V\"axj\"o.

Fuchs, C., 2002. The anti-V\"axj\"o interpretation of quantum mechanics.
Proc. Int. Conf. Quantum Theory: Reconsideration
of Foundations. ed. A. Yu. Khrennikov, Ser. Math. Modelling 2,
99-116, V\"axj\"o Univ. Press,  V\"axj\"o;  http://www.msi.vxu.se/forskn/quantum.pdf

Gudder, S. P., 2001. An approach to quantum probability. Proc. Conf.
Foundations of Probability and Physics, ed. A. Yu. Khrennikov.
Q. Prob. White Noise Anal. 13, 147-160, WSP, Singapore.

Hameroff,  S., 1994. Quantum coherence in microtubules.
A neural basis for emergent consciousness?  J. of Consciousness Studies,
1, 91-118.

Hameroff,   S., 1998. Quantum computing in brain microtubules? The Penrose-Hameroff Orch Or model
of consciousness. Phil. Tr. Royal Sc., London A, 1-28.

Healey, R., 1984. How many worlds? Nous 18, 591-616.

Hiley,  B.,  Pylkk\"anen, P., 1997.  Active information and cognitive science --
A reply to Kiesepp\"a. In: Brain, mind and physics.  Editors: Pylkk\"anen, P., Pylkk\"o, P., Hautam\"aki, A.
IOS Press, Amsterdam.

Hiley, B., 2000. Non-commutavive geometry, the Bohm interpretation and the mind-matter relationship.
Proc. CASYS 2000, Liege, Belgium.

Holevo, A. S., 2001. Statistical structure of quantum theory. Springer,
Berlin-Heidelberg.

Hopfield,  J. J., 1982. Neural networks and physical systems with
emergent collective computational abilities. Proc. Natl. Acad. Sci. USA
 79,  1554-2558.

Hoppensteadt, F. C., 1997.  An introduction to the mathematics of neurons:
modeling in the frequency domain.  Cambridge Univ. Press,
New York.

Jibu, M., Yasue,  K., 1992. A physical picture of Umezawa's quantum brain dynamics.
In  Cybernetics and Systems Research, ed. R. Trappl, World Sc., London.

Jibu, M.,  Yasue,  K., 1994.  Quantum brain dynamics and consciousness.
J. Benjamins Publ. Company, Amsterdam/Philadelphia.

Khrennikov, A. Yu., 1999. Classical and quantum mechanics on information spaces
with applications to cognitive, psychological,
social and anomalous phenomena. Found. Phys. 29,  1065-1098.

Khrennikov,  A. Yu., 2000.  Classical and quantum mechanics on $p$-adic trees of ideas.
BioSystems   56, 95-120.

Khrennikov, A. Yu., 2002. On foundations of quantum theory.
Proc. Conf. Quantum Theory: Reconsideration
of Foundations, ed. A. Yu. Khrennikov.
Ser. Math. Modelling 2, 163-196, V\"axj\"o Univ. Press,  V\"axj\"o.

Khrennikov, A. Yu., 2002.  V\"axj\"o interpretation of quantum mechanics,
http://xxx.lanl.gov/abs/quant-ph/0202107.

Khrennikov,  A. Yu., 2001. Linear representations of probabilistic
transformations induced by context transitions. J. Phys. A: Math. Gen.
34, 9965-9981.

Khrennikov, A. Yu., 2003. Contextual viewpoint to quantum stochastics.
J. Math. Phys.  44, N. 6, 2471- 2478.

Khrennikov, A. Yu., 2003. Representation of the Kolmogorov model having all distinguishing
features of quantum probabilistic model.  Phys. Lett. A 316, 279-296.

Khrennikov, A. Yu., 2002. On cognitive experiments to test quantum-like behaviour
of mind. Rep. V\"axj\"o Univ.: Math. Nat. Sc. Tech., N 7;
http://xxx.lanl.gov/abs/quant-ph/0205092.

Khrennikov, A. Yu., 2003. Hyperbolic quantum mechanics. Advances in Applied Clifford Algebras
13(1),  1-9.

Khrennikov, A. Yu., 2003. Interference of probabilities and number field structure of quantum models.
Annalen  der Physik 12, N. 10, 575-585.

Khrennikov,  A. Yu., 2003. Quantum-like formalism for cognitive measurements. Biosystems
70, 211-233.

Khrennikov,  A. Yu., 2003. Schr\"odinger dynamics as the Hilbert space projection of  a
realistic contextual probabilistic dynamics. Europhysics Letters, accepted to publication;
http://xxx.lanl.gov/abs/quant-ph/0409128.

Lockwood, M., 1989. Mind, Brain and Quantum. Oxford, Blackwell.

Lockwood, M., 1996. Many minds interpretations of quantum mechanics. British J. for the Philosophy of Sc.
47/2, 159-88.

Loewer, B., 1996. Comment on Lockwood. British J. for the Philosophy of Sc.
47/2, 229-232.

Mackey,  G. W., 1963.  Mathematical foundations of quantum mechanics.
W. A. Benjamin INc, New York.

Orlov, Y. F., 1982.  The wave logic of consciousness: A hypothesis.
Int. J. Theor. Phys. 21, N 1, 37-53.

Plotnitsky, A., 2002. Quantum atomicity and
quantum information: Bohr, Heisenberg, and quantum mechanics as an
information theory, Proc. Conf.  Quantum theory:
reconsideration of foundations, ed: A. Yu. Khrennikov, Ser. Math. Modelling
 2, 309-343, V\"axj\"o Univ. Press,  V\"axj\"o.

Plotnitsky, A., 2001. Reading Bohr:
Complementarity, Epistemology, Entanglement, and Decoherence.
Proc. NATO Workshop Decoherence and its Implications for Quantum
Computations, Eds. A.Gonis and P.Turchi, p.3--37, IOS Press,
Amsterdam.

Plotnitsky, A., 2002. The spirit and the letter of Copenhagen: a response
to Andrei Khrennikov, http://xxx.lanl.gov/abs/quant-ph/0206026.

Penrose, R., 1989. The emperor's new mind. Oxford Univ. Press, New-York.

Penrose, R., 1994.  Shadows of the mind. Oxford Univ. Press, Oxford.

Shiryayev,  A. N., 1991. Probability. Springer, Heidelberg.

Stapp, H. P.,  1993. Mind, matter and quantum mechanics.  Springer-Verlag,
Berlin-New York-Heidelberg.

Strogatz,  S. H., 1994.  Nonlinear dynamics and chaos with applications to physics,
biology, chemistry, and engineering. Addison Wesley, Reading, Mass.

van Gelder, T., Port, R., 1995. It's about time: Overview of the dynamical approach to cognition.
in  Mind as motion: Explorations in the dynamics of cognition.
Ed.: T. van Gelder, R. Port. MITP, Cambridge, Mass, 1-43.

van Gelder, T., 1995. What might cognition be, if not computation?
J. of Philosophy 91, 345-381.

Vitiello, G., 2001.  My double unveiled - the dissipative quantum model of brain.
J. Benjamins Publ. Company, Amsterdam/Philadelphia.

von Neumann, J.,  1955. Mathematical foundations
of quantum mechanics. Princeton Univ. Press, Princeton, N.J..

Whitehead, A. N., 1929. Process and Reality: An Essay in Cosmology.  Macmillan Publishing Company, New York.

Whitehead, A. N., 1933.  Adventures of Ideas.  Cambridge Univ. Press, London.

Whitehead, A. N., 1939. Science in the modern world. Penguin, London.

\end{document}